\documentclass{aastex}
\usepackage{spr-astr-addons}
\usepackage{url}
\usepackage{amsmath}
\usepackage{amsfonts}
\usepackage{amssymb}
\usepackage{graphicx}
\usepackage{subfigure}

\RequirePackage{color}

\begin{document}

\title{ Holographic dark energy with time depend gravitational constant in the non-flat  Ho$\check{r}$ava-Lifshitz cosmology}
\slugcomment{Not to appear in Nonlearned J., 45.}
\shorttitle{Short article title}

\shortauthors{Kh. Saaidi et al.}
\author{$^{a,b}$Kh. Saaidi\altaffilmark{1}} \and \author{$^b$A. aghamohammadi\altaffilmark{2}}\and \author{ $^a$M. R. Setare\altaffilmark{3}}
\affil{$^a$Department of Physics, Faculty of Science, University of
Kurdistan,  Sanandaj, Iran.} \and \affil{$^b$Faculty of Science,  Islamic Azad University - Sanandaj Branch, Iran }




\altaffiltext{1}{ksaaidi@uok.ac.ir.}
\altaffiltext{2}{agha35484@yahoo.com.}
\altaffiltext{3}{ rezakord@ipm.ir}

\begin{abstract}
We study  the  holographic dark energy on the subject of Ho$\check{r}$ava-Lifshitz gravity with a time dependent gravitational constant (G(t)),   in the non-flat  space-time. We obtain the differential equation that specify the evolution of the dark energy density parameter based on varying gravitational constant. we find out a relation  for the state parameter  of the dark energy equation of state  to low redshifts which containing varying $G$ corrections in the non-flat  space-time.
\end{abstract}

\keywords{Dark energy; future event horizen; Ho$\check{r}$ava-Lifshitz; gravitation constant; }

\section{Introductions}
The observation that  universe appears to be accelerating currently has caused one of the important probes
 to modern cosmology.  Observational data such as, the type Ia (\citeauthor{1}), supernovas cosmic microwave background
 (\citeauthor{2}), and large-scale structure (\citeauthor{3}), indicates that the accelerated expansion of the universe is
  not proceeding as that  predicted by general relativity. A first direction that could
provide an explanation of this remarkable phenomenon is to
introduce the concept of dark energy, with the most obvious
theoretical candidate being the cosmological constant. However, at
least in an effective level, the dynamical nature of dark energy
can also originate from various fields, such is a canonical scalar
field (quintessence) (\citeauthor{quint}), a phantom field, that is a
scalar field with a negative sign of the kinetic term
(\citeauthor{phant}), or the combination of quintessence and phantom in a
unified model named quintom (\citeauthor{quintom}). The second direction
that could explain the acceleration is to modify the gravitational
theory itself, such is the generalization to $f(R)$-gravity
(\citeauthor{12}), scalar tensor theories with non-minimal coupling
(\citeauthor{9}) etc.\\
 Although, there isn't a quantum theory of gravity, one can proceed to investigate the nature of dark energy based on some principles of quantum gravity.
 Currently, an interesting
such an attempt is the so-called ``Holographic Dark Energy'' (HDE)
proposal (\citeauthor{Hsu:2004ri}). The HDE is defined by
 \begin{equation} \label{1e}
\rho_{\Lambda}= 3c^2 M_p^2L^{-2},
\end{equation}
where $c^2$ is a numerical constant of order unity and $M_p$ is denoted the reduced planck mass
$M_p^{-2}=8\pi G$. The holographic dark energy scenario has been tested and
constrained by various astronomical observations (\citeauthor{obs3a}) and
it has been extended to various frameworks
(\citeauthor{nonflat,holoext,intde}).\\
 There are significant indications that Newton's
``constant'' $G$ can by varying, being a function of time or
equivalently of the scale factor (\citeauthor{G4com}). In particular,
observations of Hulse-Taylor binary pulsar (\citeauthor{Damour,kogan}),
helio-seismological data (\citeauthor{guenther}), and astereoseismological data from the
pulsating white dwarf star G117-B15A (\citeauthor{Biesiada}) lead to
$\left|\dot{G}/G\right| \lessapprox 4.10 \times 10^{-11} yr^{-1}$,
for $z\lesssim3.5$ (\citeauthor{ray1}). Thus, in previous paper
(\citeauthor{setare}), we investigated the holographic dark energy
scenario under a varying gravitational constant in the framework of Horava-Lifshitz gravity and we extracted
the corresponding corrections to the dark energy equation-of-state
parameter.\\
Recently, a power-counting renormalizable, ultra-violet (UV)
complete theory of gravity was proposed by Ho\v{r}ava in
(\citeauthor{hor2},\citeauthor{hor1},\citeauthor{hor4}). Although presenting an infrared (IR)
fixed point, namely General Relativity, in the  UV the theory
possesses a fixed point with an anisotropic, Lifshitz scaling
between time and space of the form $x^{i}\to\ell~x^{i}$,
$t\to\ell^z~t$, where $\ell$, $z$, $x^{i}$ and $t$ are the scaling
factor, dynamical critical exponent, spatial coordinates and
temporal coordinate, respectively.\\
n the present work we are interested to study the
 Holographic dark energy in framework of Ho\v{r}ava-Lifshitz
 gravity. We extend previous study (\citeauthor{setare}) to the non-flat case and we want consider the effect of curvature constant of non flat space  on the resalts which is obtained in the flat space, it is seen that the resonable range of  $\lambda$    Eq.(\ref{2g}) in non flat space is larger than answer range of $\lambda$ in flat space . The paper is organized as follows: in the next section we found  the holographic dark energy with gravitation constant depend on time and derive the differential equation that specify the evolution of dark energy parameter. In Sec. 3,  we obtain the parameter of the dark energy equation of state at the low redshift. Eventually, the latter section is devoted to conclusion.
\section{Holographic Dark Energy With Gravitational Constant Depend on Time in a  non-flat Universe}
In the case where the space-time geometry is a non-flat
Robertson-Walker:
\begin{equation}\label{1g}
ds^2= dt^2 -a^2(t)\left(\frac{1}{1-kr^2}dr^2+r^2d^2\Omega \right).
\end{equation}
with $a(t)$ the scale factor, in comoving coordinates  $(t,r,\theta,\varphi)$, where $k$ denotes
the spacial curvature with $k=-1,0,1$ corresponding to open, flat
and closed universe respectively. In this case, the first
Friedmann equation in the framework of Ho\v{r}ava-Lifshitz
 gravity writes (\citeauthor{setare}):
\begin{eqnarray}\label{2g}
H^2=\frac{\kappa^2}{6\left( 3\lambda-1\right) }\rho -\frac{\beta k}{a^2},
\end{eqnarray}
where, $\rho=\rho_m+\rho_{\Lambda}$ is the energy density, $\rho_m=\rho_{m_0} a^{-3}$ and $\rho_{\Lambda}$ are dark matter density and  dark energy density respectively, $\lambda$ is a dimensional constant and $\beta=\frac{\kappa^4\mu^2\Lambda }{8(3\lambda-1)^2}$. Here $\kappa^2=8\pi G$, and $\rho_{m_0}$ indicate the present value that quantity.    Then, from   the Eq.(\ref{2g}), we introduce the effective density parameter $\Omega_{\Lambda_e}=\frac{\Omega_{\Lambda}}{2\left( 3\lambda-1\right)}\equiv \frac{8\pi G}{6\left( 3\lambda-1\right) }\frac{\rho_{\Lambda}}{H^2}$. Substituting  the Eq.(\ref{1e}) into $\Omega_{\Lambda}$ we obtain:
\begin{eqnarray}\label{3g}
\Omega_{\Lambda e}=\frac{c_e^2}{H^2L^2},
\end{eqnarray}
where $c_e=c/\sqrt{2\left(3\lambda-1 \right)} $.
In this case, the cosmological length $L$ in (\ref{3g}) is
considered to be  (\citeauthor{nonflat}):
\begin{equation}\label{g4}
L\equiv\frac{a(t)}{\sqrt{|k|}}\,\text{sinn}\left(\sqrt{|k|}y\right), \quad y=\frac{d_h}{a(t)}
\end{equation}
where
\begin{equation}\frac{1}{\sqrt{|k|}}\text{sinn}(\sqrt{|k|}y)=
\begin{cases} \sin y  & \, \,k=+1,\\
             y & \, \,  k=0,\\
             \sinh y & \, \,k=-1.\\
\end{cases}
\end{equation}
and $d_{h}$ is the future event horizon defind by
\begin{eqnarray}\label{4g}
 d_h(a)=a\int_{t}^{\infty}\frac{dt'}{a(t')}=a\int_{a}^{\infty}\frac{da'}{Ha'^2}.
\end{eqnarray}
Henceforth, we will use $\ln a$, as an independent variable. Therefore, we define, $\dot{X}=\frac{dX}{dt}$,  and  $X'=\frac{dX}{d\ln a}$, so that $\dot{X}=X'H$. A straightforward calculation using (\ref{3g}) and
(\ref{2g})
 leads to (\citeauthor{Jamil:2009sq}):
   \begin{eqnarray}\label{5g}
 \frac{\Omega'_{\Lambda _e}}{\Omega_{\Lambda_ e}}=2 \left(\frac{\sqrt{\Omega_{\Lambda_e}}}{c_e}\text{cosn} (\sqrt{\vert k\vert}y)-1-\frac{\dot{H}}{H^2} \right).
\end{eqnarray}
where
\begin{equation}\text{cosn}(\sqrt{|k|}y)=
\begin{cases} \cos y  & \, \,k=+1,\\
             1 & \, \,  k=0,\\
             \cosh y & \, \,k=-1.\\
\end{cases}\end{equation}
In order to clarify  the effect of variety G on the $\Omega_{\Lambda e}$, we should to get rid of $\dot{H}$ into Eq. (\ref{5g}). In this regard, differentiation of the Friedmann equation give rise to
\begin{equation}\label{6g}
\dot{H}=\frac{2\pi}{3\left( 3\lambda-1\right) }\left[G'-3G\left( 1+\omega\right)  \right]\rho+\frac{\beta k}{a^2},
\end{equation}
where, we using from the fluid equation,  $\dot{\rho}=-3H(1+\omega)\rho$. Since, $\omega$ is
\begin{equation}\label{7g}
\omega=\frac{\omega_{\Lambda}\rho_{\Lambda}}{\rho}=\frac{\omega_{\Lambda}\Omega_{\Lambda _e}}{\Omega_{m_e} +\Omega_{\Lambda_ e}}.
\end{equation}
  The Eos parameter for HDE is given by
  \begin{equation}\label{8g}
\omega_{\Lambda}=-\left( \frac{1}{3}+\frac{2\sqrt{\Omega_{\Lambda}}}{3c}\right).
\end{equation}
Substituting the Eqs. (\ref{7g}), (\ref{8g}) into the Eq. (\ref{6g}) gives
\begin{eqnarray}\label{9g}
\frac{\dot{H}}{H^2}&=&\frac{{\cal G}\left( \Omega_{m_e}+\Omega_{\Lambda_ e}\right) }{2}  -\frac{3\left( \Omega_{m_e}+\Omega_{\Lambda_ e}\right) }{2}\cr
&+&\frac{\Omega_{\Lambda _e}}{2}+\frac{{ \Omega_{\Lambda_e}}^{3/2}}{c_e}-\Omega_{k_e},
\end{eqnarray}
where $\Omega_{ke}=\beta \Omega_k=-\beta k/a^2H^2,$ is the curvature parameter and ${\cal G}=G'/G$
\begin{eqnarray}\label{10g}
\frac{\Omega'_{\Lambda }}{{\Omega_{\Lambda}}}&=&\frac{2\sqrt{\Omega_{\Lambda _e}}}{c_e}\text{cosn} (\sqrt{\vert k\vert}y)+(1-\Omega_{ke})(1-{\cal G})-\Omega_{\Lambda_e}\cr
&-&\frac{2{\Omega}^{3/2}_{\Lambda_e}}{c_e}
.
\end{eqnarray}
\section{Some Cosmology Application}
Here, we should follow up an expression related to the state parameter of  equation at the present time. Since we have extracted the expressions for $\Omega_{\Lambda}'$, we can
calculate $w(z)$ for small redshifts $z$, performing the standard
expansions of the literature. In particular, since $\rho_\Lambda \sim
a^{-3(1+w)}$ we acquire
 Expanding $\rho_{\Lambda}$  we have:
  \begin{equation}\label{1s}
  \ln \rho_{\Lambda}=\ln \rho^0_{\Lambda}+ \frac{d\ln \rho_{\Lambda}}{d \ln a}\ln a
+\frac{1}{2}\frac{d^2\ln \rho_{\Lambda}}{d{(\ln a)}^2}{(\ln a)}^2+\cdots ,
  \end{equation}
here, the derivatives are taken at the present time $a_0=1$. Then, $w(z)$ is given in the small  red shifts $\ln a=-\ln (1+z)=-z$ up to second order,  as:
 \begin{equation}\label{2s}
\omega(z)=-1-\frac{1}{3}\left( \frac{d\ln \rho_{\Lambda}}{d \ln a}
-\frac{1}{2}\frac{d^2\ln \rho_{\Lambda}}{d{(\ln a)}^2}{(z)}\right).
\end{equation}
We can rewrite (\ref{2s}) as:
\begin{eqnarray}\label{4s}
\omega=\omega_0+\omega_1 z.
\end{eqnarray}
 Since
\begin{equation}\label{5s}
\rho_{\Lambda}=\frac{3\left(3\lambda-1 \right)H^2\Omega_{\Lambda_e} }{4\pi G}=\frac{\Omega_{\Lambda _e}\rho_m}{\Omega_{m_e}}=\frac{\rho_{m_0}\Omega_{\Lambda e}a^{-3}}{1-\Omega_{\Lambda_ e}-\Omega_{k_e}},
\end{equation}
after some calculation, and  some simplification we achieve to $\omega_0,  \omega_1$ as follows:
\begin{equation}\label{6s}
\omega_0=-\frac{1}{3}\frac{\left[  \Omega'_{k_e}\Omega_{\Lambda_ e}+\left(1-\Omega_{k_e} \right)\Omega'_{\Lambda_ e}\right]  }{\Omega_{\Lambda_ e}\left(1-\Omega_{\Lambda_ e}-\Omega_{k_e} \right) }
\end{equation}
\begin{eqnarray}\label{7s}
\omega_1&=&\frac{1}{6}\left[ \left(\frac{\Omega'_{\Lambda_e}}{\Omega_{\Lambda_e}} \right)'+\frac{\Omega''_{\Lambda_e}+\Omega''_{k_e}}{\left(1-\Omega_{\Lambda_e}-\Omega_{k_e} \right) }\right]\cr
& +&\frac{1}{6}\frac{\left( \Omega'_{\Lambda_e}+\Omega'{k_e}\right)^2 }{\left( 1-\Omega_{\Lambda_e}-\Omega_{k_e}\right)^2 }
\end{eqnarray}
 Now, substituting  Eq.(\ref{10g}) into Eqs.(\ref{6s}), (\ref{7s}) we obtain:
 \begin{eqnarray}\label{8s}
\omega_0&= &\frac{-1}{3}\left( \frac{{\Omega}'_{k_e}+\left( 1-\Omega_{k_e}\right)\chi }{1-\Omega_{k_e}-\Omega_{\Lambda_ e}}\right)
\end{eqnarray}
  \begin{eqnarray}\label{9s}
\omega_1&=&\frac{1}{6}\left[\frac{\left( 1-\Omega_{k_e}\right)\left({\chi}^2+\Omega_{\Lambda_e}\zeta\chi-\xi \right)  }{1-\Omega_{k_e}-\Omega_{ke}} \right]\cr
&+&\frac{1}{6}{\biggl[}{-\chi}^2+\frac{{\Omega_{Ke}}''}{1-\Omega_{\Lambda_e}-\Omega_{ke}}
{\biggl]}\cr
&+&\frac{1}{6}\frac{{\Omega}^2_{\Lambda_e}{\chi}^2+2\Omega_{\Lambda_e}\chi{\Omega}'_{ke}}{{(1-\Omega_{\Lambda_e}-\Omega_{ke})}^2}
\end{eqnarray}
where, $\chi, \zeta, \xi$ are define as follows:
\begin{eqnarray}
\chi&=&2\frac{\Omega_{\Lambda_e}}{c_e}\text{cosn}(\sqrt{|k|}y)+(1-\Omega_{ke})(1-{\cal G})\cr
&-&\Omega_{\Lambda_e}-\frac{2{{\Omega_{\Lambda_e}}^{3/2}}}{c_e},\cr
\zeta&=&\frac{1}{c_e\sqrt{\Omega_{\Lambda_e}}}\text{cosn}(\sqrt{|k|}y)-1-\frac{3{\Omega_{\Lambda_e}}^{1/2}}{c_e},\\
\xi&=&+{\Omega_{ke}}'(1-{\cal G})+{{\cal G}'}(1-\Omega_{ke})\cr
&-&\frac{2\sqrt{\Omega_{\Lambda_e}|k|}}{a H c_e}\text{sinn}(\sqrt{|k|}y ).\nonumber
\end{eqnarray}
In the following, we want compare the diagram of  the  state parameter  equation at the present time  versus $\lambda$ on   both   flat space and non-flat space for different  redshifts. In the Fig.1 we have  plotted the equation of state parameter $\omega$, versus $\lambda$ on both flat and non flat space time for $z=0.01$ and ${\cal G}=0.2$ and small $\Omega_{ke}$. We have obtained interval $\Delta\lambda_1=(0.32, 0.93)$ and   $\Delta\lambda_2=(0.24, 1.132)$  for flat space time $(a)$and non flat space time $(b)$ respectivly, in which $\omega$  accept the allows values between $(-1, 1)$. as well as this action in the Fig.2 and Fig.3, also  have  plotted for $z=0.5$, and $z=0.9$  on  both flat (a) and non flat (b)  space time respectivly. It is clearly seen that is $\Delta\lambda_2>\Delta \lambda_1$. This result  show that the suitable  interval for $\lambda$  which $\omega$ accept the values between $(-1, 1)$ in non flat space time is larger than the relevant interval with  flat space time.
\begin{figure}[h]
\begin{minipage}[b]{1\textwidth}
\subfigure[\label{fig1a} ]{ \includegraphics[width=.2\textwidth]%
{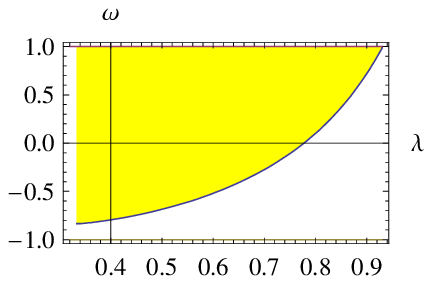}} \hspace{.2cm}
\subfigure[\label{fig1b} ]{ \includegraphics[width=.2\textwidth]%
{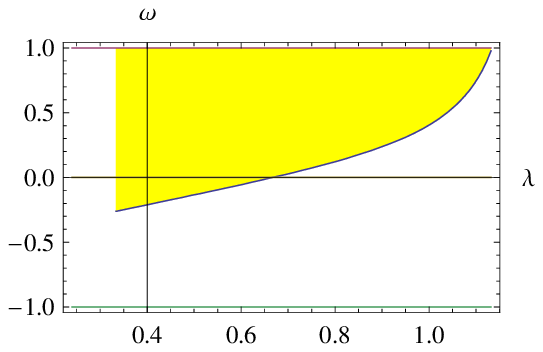}}
\end{minipage}
\caption{ the diagram Variations  the state parameter, $\,\omega$ with respect to
$\lambda$:(a) for flat space tim.  (b)non flat space time.}
\end{figure}\\
\begin{figure}[h]
\begin{minipage}[b]{1\textwidth}
\subfigure[\label{fig1a} ]{ \includegraphics[width=.2\textwidth]%
{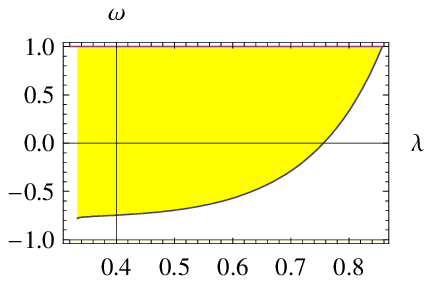}} \hspace{.2cm}
\subfigure[\label{fig1b} ]{ \includegraphics[width=.2\textwidth]%
{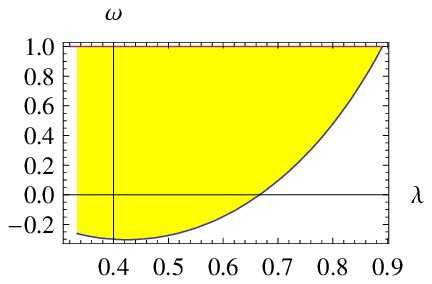}}
\end{minipage}
\caption{ the diagram Variations  the state parameter, $\,\omega$ with respect to
$\lambda$ for  $z=0.5$ : (a)   for flat space time.It is seen that the interval $(0.32 , 0.857)$ is satisfyed  for $\lambda$ in which $\omega$ accept the allows values between $-1, 1$. (b) for non flat space time. .It is seen that
interval $(0.32 , 0.89)$ is satisfyed for $\lambda$ to the same interval of $\omega$.  }
\end{figure}\\
\begin{figure}[h]
\begin{minipage}[b]{1\textwidth}
\subfigure[\label{fig1a} ]{ \includegraphics[width=.2\textwidth]%
{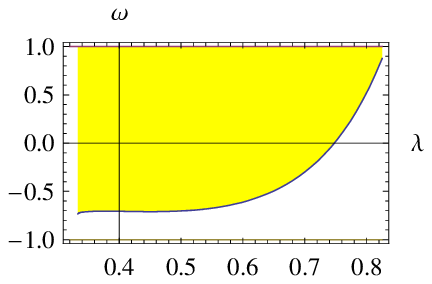}} \hspace{.2cm}
\subfigure[\label{fig1b} ]{ \includegraphics[width=.2\textwidth]%
{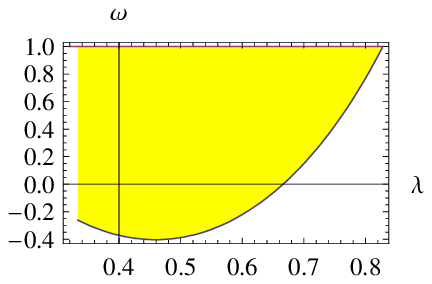}}
\end{minipage}
\caption{ the diagram Variations  the state parameter, $\,\omega$ with respect to
$\lambda$ for $z=0.9$: (a)  for flat space time. It is seen that the interval $(0.32 , 0.825)$is satisfyed  for $\lambda$ in which $\omega$ accept the allows values between $-1, 1$. (b)on non flat space. It is seen that the
interval $(0.32 , 0.827)$ is satisfyed for $\lambda$ in which $\omega$ accept the allows values between $(-1, 1)$,  }
\end{figure}
\newpage
\section{Conclusion}
Astrophysical observations imply that the  state parameter of dark energy must be changeable on the cosmic time. In this regard, one obvious contender is the holographic dark energy. In the HDE, the parameter $c$
can take  the various values,  but we set $c=1$.     In this paper we have discussed  the holographic dark energy whit  the  time dependent gravitation constant  in the framework of Ho$\check{r}ava$ gravity in the non flat universe. By evaluating  Eqs. (\ref{8s}), (\ref{9s}), and their diagram,  it is considerable that anyone from values  of limeted $\lambda$  gives a suitable estimate from the state parameter, which is agreeable with experience data.
Choosing  $\Omega_{\Lambda}=0.73$, for the average vlue  $0.32<\lambda<0.84$, one can attain to $-0.4<\omega<1$ in flat and nonflat space time and the only effect of  the curvature constant is that the interval of the $\lambda$ variations  outgrow lightly from  that one   .  However,  by applying the term curvature and the $cos$ term due to curvature parameter, although variation  of $\omega$ isn't notable, nevertheless these could be significant in the overall of cosmological evaluation.
The last statement is that, however, the
effect of variation $k$ is very small, but it has an important role  in the evolution of cosmological and it is shown that the equation of state of dark energy for lower red shift is generalized due to contribution of applying $k$.

\end{document}